# Joint probability density of a passive article with force and magnetic field

**Jae-Won Jung[1], Sung Kyu Seo[1], Kyungsik Kim[1, 2]**

[1] DigiQuay Company Ltd., Seocho-gu Seoul 06552, Republic of Korea
[2] Department of Physics, Pukyong National University, Busan 608-737, Republic of Korea

E-mail: kskim@pknu.a.ckr

## Abstract

We firstly study the Navier-Stokes equation for the motion of a passive particle with harmonic, viscous, perturbative forces, subject to an exponentially correlated Gaussian force, in $t << \tau$ , $t << \tau$ and for $\tau = 0$ , where $\tau$ is the correlation time. The mean squared displacement and mean squared velocity for a passive particle with both harmonic and viscous force and perturbative force behavior like the super-diffusion scales to $\sim t^4$ in $\tau = 0$ , while those to $\sim t^3$ in $t >> \tau$ . The mean squared velocity for a particle with no harmonic and viscous forces get the Gaussian form, which is the normal diffusion with $< v^2(t) > \sim t$ in $\tau = 0$ . From moment equation, the moment for a passive particle with a harmonic and viscous forces grows with time $\mu_{2,2} \sim t^5$, while the moment with perturbative force scales to $\sim t^6$ in three-time domains. Secondly, from the Fokker-Planck equation in an incompressible conducting fluid of magnetic field, we approximately obtain the solution of the joint probability density by using double Fourier transforms in three-time domains. In the hydro-magnetic case of velocity and time-dependent magnetic field, the mean squared velocity for joint probability density of velocity and magnetic field has secondly super-diffusive form scales to $\sim t^3$ in $t >> \tau$ , while the mean squared displacement for joint probability density of velocity and magnetic field reduces to be time $\sim t^4$ in $t << \tau$ . The solution in $t >> \tau$ and $\tau = 0$ behaviors normal diffusion with the mean squared magnetic field $< h^2(t) > \sim t$ . Particularly, in short-time domain $t << \tau$ , the moment for an incompressible conducting fluid of magnetic field becomes super-diffusion with $\mu^h_{2,0,2} \sim t^6$ , consistent with our result. In addition, the kurtosis, the correlation coefficient, and the moment from moment equation are numerically calculated.

---

## 1. Introduction

The fluid motion has been descrbed and extended by Navier-Stokes equation with viscosity and pressure terms in the past time. The difference point between Navier-Stokes equation and the closely related Euler equation is that former takes into account the viscous flow, while latter the inviscid flow. Navier-Stokes equation has attracted much interest in various scientific systems, and examples of such system are utilized to model and analyze the weather, ocean current, water flow, pipe flow, and air flow. In addition, Navier-Stokes equation in its full and simplified form is applied from the microscopic fluidity study of blood flow [1, 2] and analysis of convective or diffusing pollution to the macroscopic design of power stations, modelling of magneto-hydrodynamics, design of aircraft and cars, and many other problems. Examples of models are recently included the wall-modeled large-eddy simulations [3, 4], the fluid-particle coupling method [5] including methodologies including cut-cell methods and ghost-cell methods [6-8],

On the other hand, the formulation in theory of turbulence in compressible fluids [9] has involved plausibility and approximation in the beginning. Wyld [10] has formulated the theory of turbulence introducing method of systematic perturbation similar to quantum field theory. Lee [11] has described the generalization of Wyld's formulation to the hydromagnetic equations in theory of stationary, homogeneous, isotropic turbulence in compressible fluids and the solution for velocity and magnetic field in the form of perturbation series. Forster, Nelson, and Stephen [12, 13] have argued that the renormalization group theory apply and analyze to the large-displacement, long-time behavior of velocity correlations generalized by the Navier-Stokes equation for a randomly incompressible fluid. Particularly, their results have applied a forced Burger's equation in one dimension, even though the physical understanding of Navier-Stokes equation below two dimension is not clear. As the popular and well-known models, Antonov et al. [14, 15] have coupled the Kazantsev-Kraichnan velocity ensemble describing the environment to three different models such as Kardar-Parisi-Zhang (including Navier-Stokes equation) model, Hwa-Kardar model, Pavlik mode. They have showed the significant effect leading to inducing non-linearity or making the anisotropic scaling via field theoretical renormalization group analysis. Meanwhile Navier-Stokes turbulence [16] has been of interest in two-dimensional fluid for many years. Examples in two-dimensional Navier-Stokes turbulence are the soap film flow [17, 18], rotating fluid [19], magnetically forced stratified fluid [20], two-fluid hydrodynamics [21] and plasma in the equatorial ionosphere [22]. The fractional Brownian motion [23], fractional diffusion, generalized Langevin equation, and Fokker-Planck equation [24-29] have treated and analyzed theoretically and numerically in the anomalous diffusion process and the transport process.

In this paper, we firstly derive the Fokker-Planck equation in Navier-Stokes equation for a passive particle with harmonic and viscous force, perturbative force, and no harmonic and viscous force. We approximately solve the joint probability density of displacement and velocity. Secondly, from Navier-Stokes equation for a passive particle with harmonic and viscous forces, we approximately solve the joint probability density of velocity and time-dependent magnetic field. The paper is structured as follows: In Section 2-1, we derive the Fokker-Planck equation, and introduce Fourier transform of joint probability density in order to approximately find the joint probability density and its moment. In Section 2-1, 2-2, and 2-3, we find and calculate the joint probability density and its moment in three-time domains of correlation time. Sections 2-5 presents the moment equation related displacement and velocity. In Section 2-6, we summarize the key findings. In Section 3, we approximately solve the joint probability density of velocity and time-dependent magnetic field by using a similar method of Section 2. In Section 4, we provide and account for a conclusion summarizing our key findings.

## 2. Navier-Stokes equation with harmonic and perturbative forces

### *2-1 Fokker-Planck equation in displacement and velocity*

First of all, we introduce Navier–Stokes momentum equation of modified convective form with uniform shear and bulk viscosity. Navier-Stokes equation with fluctuating force is expressed as follows [16]:

$$\frac{d}{dt}v(t) = -\frac{1}{\rho}v(t)\cdot\nabla v(t) - \frac{1}{\rho}\nabla p + \nu_0\nabla^2 v(t) - \nu_1 v(t) + f(t) + g(t) \qquad (2\text{-}1)$$

Here, $\nu_0$ is the drag coefficient, $\nu_1$ the viscosity, and $f$ an external force acting on the particle. We assume that the incompressibility condition is $\nabla\cdot v(t) = 0$, and put that the pressure term is $\frac{1}{\rho}\nabla p = 0$. Some researchers have considered Eq. (2-1) as a model of homogeneous, isotropic turbulence [16, 17]. It is well known from the absence of drag that an energy cascades to larger scales [30, 31]. In one dimension, we replace a novel Eq. (2-1) as

$$\frac{d}{dt}v(t) = -v(t)\nabla v(t) - rv(t) + e\nabla^2 v(t) - \beta x + g(t) \quad (2\text{-}2)$$

Here $f = -\beta x$ denotes a harmonic force, $\rho = 1$ the dimensionless value, and the viscosity coefficient $v_1$ replaces to $r$. The fluctuation term $g(t)$ depends exponentially on the time difference

$$< g(t) >= 0 , < g(t)g(t') >= g_0^2 g(t-t') = \frac{1}{2\tau}\exp(-\frac{|t-t'|}{\tau}) . \quad (2\text{-}3)$$

Here $g_0^2 = 2rk_B T$, $\tau$ denotes the characteristic time, the parameter $g$ coupling strength, $k_B$ Boltzmann constant, $T$ temperature, and $r$ the frictional constant. By taking time derivatives of the joint probability density, the joint probability density $P(x(t),v(t),t)$ for the displacement $x$ and the velocity $v$ is defined by $P(x(t),v(t),t) \; =< \delta(x - x(t))\delta(v - v(t)) >$. We assume in joint probability density that the particle is at rest at time $t = 0$ at initial state. Inserting Eq. (2-2) into $P(x(t),v(t),t)$, we can write the Fokker-Planck equation [32] for $P(x(t),v(t),t) \equiv P(x,v,t)$ as

$$\frac{\partial}{\partial t}P(x,v,t) = -\frac{\partial}{\partial x} < \frac{\partial x}{\partial t}\delta(x - x(t))\delta(v - v(t)) >$$

$$-\frac{\partial}{\partial v} < [-v(t)\nabla v(t) - rv(t) + e\nabla^2 v(t) + g(t)]\delta(x - x(t))\delta(v - v(t)) > . \quad (2\text{-}4)$$

Then the joint probability density is derived as

$$\frac{\partial P(x,v,t)}{\partial t} = -v\frac{\partial P(x,v,t)}{\partial x} + \frac{d}{dv}v\frac{d}{dx}vP(x,v,t) - r\frac{d}{dv}vP(x,v,t) + e\frac{d}{dv}\frac{d^2}{dx^2}v(t)P(x,v,t)$$

$$+\beta x\frac{\partial P(x,v,t)}{\partial v} - ac(t)\frac{\partial^2 P(x,v,t)}{\partial x \partial v} + ab(t)\frac{\partial^2 P(x,v,t)}{\partial v^2} . \quad (2\text{-}5)$$

Here $a = \frac{g_0^2}{2m^2}$, $b(t) = 1 - \exp(-t/\tau)$ and $c(t) = (t+\tau)\exp(-t/\tau) - \tau$ .Eq. (2-5) is called the Fokker-Planck equation, as mentioned as Introduction. Some of the derivations in relation to the color noise, $b(t)$ and $c(t)$ are given Ref. [33].

In order to find the joint probability density, we introduce the double Fourier transform $P(\xi,\eta,t) = \int_{-\infty}^{+\infty}dx\int_{-\infty}^{+\infty}dt\exp(-i\xi x - i\eta v)P(x,v,t)$. By taking double Fourier transform in Eq. (2-5), the Fourier transform of joint probability density is given by

$$\frac{\partial}{\partial t}P(\xi,\eta,t) = (\xi - r\eta - e\xi^2\eta + \xi\eta\frac{\partial}{\partial \xi})\frac{\partial}{\partial \eta}P(\xi,\eta,t) - \alpha\eta\frac{d}{d\xi}P(\xi,\eta,t) + a[c(t)\xi\eta - b(t)\eta^2]P(\xi,\eta,t) . \quad (2\text{-}6)$$

*2-2. $P(x,t)$ and $P(v,t)$ in short-time domain*

In this section, we will find $P(x,t)$ and $P(v,t)$ in short-time domain $t << \tau$ . Two equations in respect with obtaining the special solutions for $\xi$, $\eta$ by variable separation is as follows:

$$\frac{\partial}{\partial t}P(\xi,t) = -\alpha\eta\frac{\partial}{\partial \xi}P(\xi,t) + \frac{a}{2}[c(t)\xi\eta - b(t)\eta^2]P(\xi,t) + AP(\xi,t) \quad (2\text{-}7)$$

$$\frac{\partial}{\partial t}P(\eta,t) = (\xi - r\eta - e\xi^2\eta + \xi\eta D_\xi)\frac{\partial}{\partial \eta}P(\eta,t) + \frac{a}{2}[c(t)\xi\eta - b(t)\eta^2]P(\eta,t) - AP(\eta,t) . \quad (2\text{-}8)$$

Here $D_\xi \equiv \partial / \partial\xi$ and $A$ denotes the separation constant. In steady state, as $\frac{\partial}{\partial t}P(\xi,t) = 0$ and $\frac{\partial}{\partial t}P(\eta,t) = 0$ , the probability density $P(\xi,t)$, $P(\eta,t)$ becomes $P^{st}(\xi,t)$ and $P^{st}(\eta,t)$ . Then we can calculate as

$$-\beta\eta\frac{d}{d\xi}P^{st}(\xi,t) + \frac{a}{2}[c(t)\xi\eta - b(t)\eta^2]P^{st}(\xi,t) + AP^{st}(\xi,t) = 0 . \quad (2\text{-}9)$$

$$(\xi - r\eta - e\xi^2\eta + \xi\eta D_\xi)\frac{\partial}{\partial\eta}P^{st}(\eta,t) + \frac{a}{2}[c(t)\xi\eta - b(t)\eta^2]P^{st}(\eta,t) - AP^{st}(\eta,t) = 0\,. \tag{2-10}$$

In the short-time domain $t << \tau$, from Eq. (2-9), Fourier transform of the probability density in the steady state is obtained as

$$P^{st}(\xi,t) = \exp[\frac{a}{2\beta\eta}[\frac{c(t)}{2}\eta\xi^2 - b(t)\eta^2\xi] + \frac{A}{\beta\eta}\xi]\,. \tag{2-11}$$

In order to find the solutions of joint functions for $\xi$ from $P(\xi,t) \equiv Q(\xi,t)P^{st}(\xi,t)$, we can obtain the distribution functions via the calculation including terms up to order $t^2/\tau^2$. That is,

$$P(\xi,t) = Q(\xi,t)P^{st}(\xi,t) = Q(\xi,t)\exp[\frac{a}{2\beta\eta}[\frac{c(t)}{2}\eta\xi^2 - b(t)\eta^2\xi]] \tag{2-12}$$

$$Q(\xi,t) = R(\xi,t)Q^{st}(\xi,t) = R(\xi,t)\exp[\frac{a}{2(\beta\eta)^2}[\frac{b'(t)}{2}\eta^2\xi^2 - \frac{c'(t)}{6}\eta\xi^3]] \tag{2-13}$$

$$R(\xi,t) = S(\xi,t)R^{st}(\xi,t) = S(\xi,t)\exp[-\frac{a}{2(\beta\eta)^3}[\frac{b''(t)}{6}\eta^2\xi^3 - \frac{c''(t)}{24}\eta\xi^4]] \tag{2-14}$$

$$S(\xi,t) = T(\xi,t)S^{st}(\xi,t) = T(\xi,t)\exp[\frac{a}{2(\beta\eta)^4}[\frac{b'''(t)}{24}\eta^2\xi^4 - \frac{c'''(t)}{120}\eta\xi^5]]\,. \tag{2-15}$$

Here, neglecting terms proportional to $1/\tau^3$ and taking the solutions as arbitrary functions of variable $t-(\xi/\beta\eta)$, the arbitrary function $T(\xi,t)$ becomes $T_\xi(\xi,t) = \Theta[t-(\xi/\beta\eta)]$. We consequently get $P(\xi,t)$ related to $T(\xi,t)$ as

$$P(\xi,t) = \Theta[t-(\xi/\beta\eta)]S^{st}(\xi,t)R^{st}(\xi,t)Q^{st}(\xi,t)P^{st}(\xi,t)\,. \tag{2-16}$$

By expanding their derivatives to second order in powers of $t/\tau$, we respectively obtain the bilinear expressions for $P(\xi,t)$, after some cancellations. That is,

$$P(\xi,t) = \exp[-\frac{at^2}{8\beta\tau^2}[1-\frac{2}{3}[\frac{t}{\tau}]^{-1}]\xi^2 - \frac{at^3}{4\tau}[1-\frac{1}{6}[\frac{t}{\tau}]]\eta\xi - \frac{a\beta t^3}{4}[1-\frac{1}{3}[\frac{t}{\tau}]]\eta^2]\,. \tag{2-17}$$

Here

$$\Theta[u] = \exp[-\frac{a\beta}{120\tau^2}\eta^2u^5 + \frac{a\beta}{48\tau}\eta^2u^4 + \frac{a\beta t}{12\tau}\eta^2u^3 - \frac{a}{4\tau}\eta^2u^2 - \frac{a\beta t}{4}\eta^2u^2]\,. \tag{2-18}$$

Using the inverse Fourier transform, $P(x,t)$ is presented by

$$P(x,t) = \frac{1}{2\pi}\int_{-\infty}^{+\infty} d\xi \exp(-i\xi x)P_\xi(\xi,t)$$

$$= [\pi\frac{at^2}{2\beta\tau^2}[1-\frac{2}{3}[\frac{t}{\tau}]^{-1}]]^{-1/2}\exp[-\frac{2\beta\tau^2x^2}{at^3}[1-\frac{2}{3}[\frac{t}{\tau}]^{-1}]^{-1}] \tag{2-19}$$

The mean squared displacement for $P(x,t)$ is given by

$$< x^2 > = \frac{at^2}{4\beta\tau^2}[1-\frac{2}{3}[\frac{t}{\tau}]^{-1}]\,. \tag{2-20}$$

From Eq. (2-10) for $\eta$, Fourier transform of the probability density in the steady state is obtained as

$$P^{st}(\eta,t) = \exp[\int\frac{a}{(\xi - r\eta - e\xi^2\eta + \xi\eta D_\xi)}[b(t)\eta^2 - c(t)\xi\eta]d\eta$$

$$= \exp[\frac{a}{\xi}[\frac{b(t)}{3}\eta^3 - \frac{c(t)}{2}\xi\eta^2] + \frac{ar}{\xi^2}[\frac{b(t)}{4}\eta^4 - \frac{c(t)}{3}\xi\eta^3] + ae[\frac{b(t)}{4}\eta^4 - \frac{c(t)}{3}\xi\eta^3] - \frac{a}{\xi}[\frac{c(t)}{3}\eta^3]]\,. \tag{2-21}$$

Here we assume that $\frac{1}{(\xi - r\eta - e\xi^2\eta + \xi\eta D_\xi)} \cong \frac{1}{\xi}[1 - r\eta\xi^{-1} - e\xi\eta + \eta D_\xi]$, and the constant values $r$ and $e$ are relatively smaller rather than one. In order to find the solutions of joint functions for $\eta$ from $P(\eta,t) = Q(\eta,t)P^{st}(\eta,t)$, we can obtain the distribution functions in the short-time domain $t << \tau$ via the calculation including terms up to order $t^2/\tau^2$. That is,

$$P(\eta,t)=Q(\eta,t)P^{st}(\eta,t) \tag{2-22}$$

$$Q(\eta,t)=R(\eta,t)Q^{st}(\eta,t)$$
$$=\exp[\frac{a}{\xi^2}[\frac{b'(t)}{12}\eta^4-\frac{c'(t)}{6}\xi\eta^3]+\frac{ar}{\xi^3}[\frac{b'(t)}{20}\eta^5-\frac{c'(t)}{12}\xi\eta^4]+\frac{ae}{\xi}[\frac{b'(t)}{20}\eta^5-\frac{c'(t)}{12}\xi\eta^4]-\frac{a}{12\xi^2}c'(t)\eta^4] \tag{2-23}$$

$$R(\eta,t)=S(\eta,t)R^{st}(\eta,t)$$
$$=\exp[\frac{a}{\xi^3}[\frac{b''(t)}{60}\eta^5-\frac{c''(t)}{24}\xi\eta^4]+\frac{ar}{\xi^4}[\frac{b''(t)}{120}\eta^6-\frac{c''(t)}{60}\xi\eta^5]+\frac{ae}{\xi^2}[\frac{b''(t)}{120}\eta^6-\frac{c''(t)}{60}\xi\eta^5]-\frac{a}{60\xi^3}c''(t)\eta^5] \tag{2-24}$$

$$S(\eta,t)=T(\eta,t)S^{st}(\eta,t)$$
$$=\exp[-\frac{a}{\xi^3}\frac{c''(t)}{120}\eta^5-\frac{ar}{\xi^4}\frac{c''(t)}{360}\eta^6-\frac{ae}{\xi^2}\frac{c''(t)}{360}\eta^6-\frac{a}{360\xi^4}c'''(t)\eta^6]\cdot \tag{2-25}$$

Neglecting terms proportional to $1/\tau^3$ and taking the solutions as arbitrary functions of variable $t+\eta/(\xi-r\eta-e\xi^2\eta+\xi\eta D_\xi)$ , the arbitrary function $T(\eta,t)$ also becomes $T(\eta,t)=\Theta[t+\eta/(\xi-r\eta-e\xi^2\eta+\xi\eta D_\xi)]$. We find that

$$P(\eta,t)=\Theta[t+\eta/(\xi-r\eta-e\xi^2\eta+\xi\eta D_\xi)]S^{st}(\eta,t)R^{st}(\eta,t)Q^{st}(\eta,t)P^{st}(\eta,t)\,. \tag{2-26}$$

By expanding their derivatives to second order in powers of $t/\tau$, we respectively obtain the bilinear expressions for $P(\eta,t)$, after some cancellations. That is,

$$P(\eta,t)=\exp[-\frac{at^4}{8\tau}[1+\frac{2}{t}]\xi^2-\frac{art^4}{4\tau}[1+\frac{1}{rt}]\eta\xi-\frac{2art^4}{3\tau}[1-\frac{3\tau}{2t}]\eta^2]\,. \tag{2-27}$$

Using the inverse Fourier transform, $P(v,t)$ is presented by

$$P(v,t)=\frac{1}{2\pi}\int_{-\infty}^{+\infty}d\eta\exp(-i\eta v)P(0,\eta,t)=[8\pi\frac{art^4}{3\tau}[1+\frac{1}{2t}]]^{-1/2}\exp[-\frac{3\tau v^2}{8art^4}[1-\frac{3\tau}{2t}]^{-1}]\cdot \tag{2-28}$$

The mean squared velocity for $P(v,t)$ is given by

$$<v^2>=\frac{4art^4}{3\tau}[1-\frac{3\tau}{2t}]\cdot \tag{2-29}$$

### *2-3 $P(x,t)$ and $P(v,t)$ in long-time domain*

In this section, we will find $P(x,t)$ and $P(v,t)$ in long-time domain $t>>\tau$. In long-time domain, we write approximate equation for $\xi$, from Eq. (2-7),

$$\frac{\partial}{\partial t}P_\xi(\xi,t)\cong\frac{a}{2}[c(t)\xi\eta-b(t)\eta^2]P_\xi(\xi,t)\cdot \tag{2-30}$$

Then we simply calculate $P_\xi(\xi,t)$ as

$$P_\xi(\xi,t)=\exp[\frac{a}{2}\int[c(t)\xi\eta-b(t)\eta^2]\,dt\cdot \tag{2-31}$$

Here $\int b(t)dt=t-\tau$, $b(t)=1$ and $\int c(t)dt=-\tau t$, $c(t)=-\tau$ in long-time domain. We find the steady probability density $Q_\xi^{st}(\xi,t)$ for $\xi$ from $P_\xi(\xi,t)\equiv Q_\xi(\xi,t)P_\xi^{st}(\xi,t)$, that is,

$$Q_\xi^{st}(\xi,t)=\exp[\frac{a}{2}\int[b(t)\eta^2-c(t)\xi\eta]\,dt]\cdot \tag{2-32}$$

From Eq. (2-9), we have the steady probability density $P^{st}(\xi,t)$ and the probability density $Q(\xi,t)$ in long-time domain $t>>\tau$ as

$$P^{st}(\xi,t)=\exp[\frac{a}{2\beta\eta}[\frac{c(t)}{2}\eta\xi^2-b(t)\eta^2\xi]] \tag{2-33}$$

$$Q(\xi,t)=R(\xi,t)Q_\xi^{st}(\xi,t)\,. \tag{2-34}$$

Taking the solutions as arbitrary functions of variable $t-\xi/\beta\eta$ , the arbitrary function $R(\xi,t)$ becomes $R(\xi,t)=\Theta[t-[\xi/\beta\eta]]$. Consequently, as expanding their derivatives to second order in powers of $t/\tau$, we obtain the expression for $P(\xi,t)$ after some cancellations, that is,

$$P(\xi,t)=R(\xi,t)Q_{\xi}^{st}(\xi,t)P^{st}(\xi,t)=\Theta[t-[\xi/\beta\eta]])Q_{\xi}^{st}(\xi,t)P^{st}(\xi,t)\cdot \quad (2\text{-}35)$$

Here $$\Theta(u)=\exp[\frac{a}{2}\beta\tau t\eta^2 u-\frac{a}{2}(t-\tau)\eta^2+\frac{a\beta\tau}{4}v^2u^2-\frac{a}{2}\eta^2 u]\cdot \quad (2\text{-}36)$$

By using the method similar to the $\xi'$ s case, we also write approximate equation for $\eta$ from Eq. (2-7) as

$$\frac{\partial}{\partial t}P_{\eta}(\eta,t)\cong a[c(t)\xi\eta-b(t)\eta^2]P_{\eta}(\eta,t)\cdot \quad (2\text{-}37)$$

We simply calculate $P_{\eta}(\eta,t)$ as

$$P_{\eta}(\eta,t)=\exp[\frac{a}{2}\int[c(t)\xi\eta-b(t)\eta^2]\,dt\cdot \quad (2\text{-}38)$$

We find $Q_{\eta}^{st}(\eta,t)$ from $P(\eta,t)=Q_{\eta}^{st}(\eta,t)P_{\eta}(\eta,t)$ as

$$Q_{\eta}^{st}(\xi,t)=\exp[\frac{a}{2}\int[b(t)\eta^2-c(t)\xi\eta]\,dt]\cdot \quad (2\text{-}39)$$

From Eq. (2-10), we have for long-time domain $t>>\tau$

$$P^{st}(\eta,t)=\exp[\frac{a}{\xi}[\frac{b(t)}{3}\eta^3-\frac{c(t)}{2}\xi\eta^2]+\frac{ar}{\xi^2}[\frac{b(t)}{4}\eta^4-\frac{c(t)}{3}\xi\eta^3]+ae[\frac{b(t)}{4}\eta^4-\frac{c(t)}{3}\xi\eta^3]-\frac{a}{\xi}[\frac{c(t)}{3}\eta^3]] \quad (2\text{-}40)$$

$$Q_{\eta}(\eta,t)=R(\eta,t)Q_{\eta}^{st}(\xi,\eta,t)=R(\eta,t)\exp[\frac{a}{2}\int[b(t)\eta^2-c(t)\eta\xi]\,dt]\cdot \quad (2\text{-}41)$$

Taking the solutions as arbitrary functions of variable $t+\eta/(\xi-r\eta-e\xi^2\eta+\xi\eta D_{\xi})$, the arbitrary function $R(\eta,t)$ becomes $R(\eta,t)=\Theta[t+\eta/(\xi-r\eta-e\xi^2\eta+\xi\eta D_{\xi})]$. Consequently, as expanding their derivatives to second order in powers of $t/\tau$, we obtain the expression for $P(\eta,t)$ after some cancellations. That is,

$$P(\eta,t)=R(\eta,t)Q_{\eta}^{st}(\xi,\eta,t)P^{st}(\eta,t)=\Theta[t+\eta/(\xi-r\eta-e\xi^2\eta+\xi\eta D_{\xi})]Q_{\eta}^{st}(\xi,\eta,t)P^{st}(\xi,\eta,t)\cdot \quad (2\text{-}42)$$

For long-time domain $t>>\tau$ , from Eq. (2-35) and Eq. (2-42), we have

$$P(\xi,\eta,t)=P(\xi,t)P(\eta,t)=\exp[-\frac{at^4}{6\tau}[1-\frac{2}{ct}]\xi^2-\frac{ar^2t^4}{2}[1-\frac{3}{rt}]\eta\xi-\frac{ar^2t^3}{2}[1-\frac{1}{rt}]\eta^2]\cdot \quad (2\text{-}43)$$

By using the inverse Fourier transform, $P(x,t)$ and $P(v,t)$ are, respectively, presented by

$$P(x,t)=[2\pi\frac{at^4}{3\tau}[1-\frac{2}{ct}]]^{-1/2}\exp[-\frac{3\tau x^2}{2at^4}[1-\frac{2}{ct}]^{-1}] \quad (2\text{-}44)$$

$$P(v,t)=[2\pi ar^2t^3[1-\frac{1}{rt}]]^{-1/2}\exp[-\frac{x^2}{2ar^2t^3}[1-\frac{1}{rt}]^{-1}]\cdot \quad (2\text{-}45)$$

The mean squared displacement and velocity from Eq. (2-44) and eq. (2-45) are

$$<x^2(t)>=\frac{at^4}{3\tau}[1-\frac{2}{ct}] \quad (2\text{-}46)$$

$$<v^2(t)>=ar^2t^3[1-\frac{1}{rt}]\cdot \quad (2\text{-}47)$$

The calculated process is complicated, but the mean squared displacement and velocity are surprisingly simple form with time $\sim t^4$ and $\sim t^3$. One find that the result is different that applied only the random force [12].

### *2-4* $P(x,t)$ *and* $P(v,t)$ *in* $\tau=0$

In this section, we will find $P(x,t)$ and $P(v,t)$ in time domain $\tau=0$ ($t\to\infty$). In $\tau=0$ domain ($b(t)=1$, $c(t)=0$), we write approximate equations from Eq. (2-9) and (2-10) for $\xi$ and $\eta$

$$\frac{\partial}{\partial t}P(\xi,t)\cong-\beta\eta\frac{\partial}{\partial\xi}P(\xi,t)-\frac{a}{2}\eta^2P(\xi,t) \quad (2\text{-}48)$$

$$\frac{\partial}{\partial t}P(\eta,t)\cong(\xi-r\eta-e\xi^2\eta+\xi\eta D_\xi)\frac{\partial}{\partial\eta}P^{st}(\eta,t)-\frac{a}{2}\eta^2P(\eta,t)\,. \quad (2\text{-}49)$$

In steady state, we can calculate $P^{st}(\xi,t)$ and $P^{st}(\eta,t)$ as

$$P^{st}(\xi,t)=\exp[-\frac{\alpha}{2\beta\eta}\eta^2\xi-\frac{A}{\beta\eta}\xi] \quad (2\text{-}50)$$

$$P^{st}(\eta,t)=\exp[\frac{a}{3\xi}\eta^3+\frac{ar}{4\xi^2}\eta^4+\frac{ae}{4}\eta^4]\,. \quad (2\text{-}51)$$

We find $P(\xi,t)$ and $P(\eta,t)$ as

$$P(\xi,t)=\Theta[t-\frac{\xi}{\beta\eta}]P^{st}(\xi,t) \quad (2\text{-}52)$$

$$P(\eta,t)=\Theta[t+\eta/(\xi-r\eta-e\xi^2\eta+\xi\eta D_\xi)]P^{st}(\eta,t)\,. \quad (2\text{-}53)$$

Thus we calculate $P(\xi,\eta,t)$ from Eq. (2-52) and (2-53) as

$$P(\xi,\eta,t)=P(\xi,t)P(\eta,t)=\exp[-\frac{art^4}{8}[1+\frac{4}{3rt}]\xi^2-\frac{art^3}{2}[1+\frac{1}{rt}]\eta\xi-\frac{3ar^3t^4}{4}[1+\frac{2}{rt}]\eta^2]\,. \quad (2\text{-}54)$$

Using the inverse Fourier transform, $P(x,t)$ and $P(v,t)$ are, respectively, presented by

$$P(x,t)=[\pi\frac{\alpha rt^4}{2}[1+\frac{4}{3rt}]]^{-1/2}\exp[-\frac{2x^2}{\alpha rt^4}[1+\frac{4}{3rt}]^{-1}] \quad (2\text{-}55)$$

$$P(v,t)=[\pi 3ar^3t^4[1+\frac{2}{rt}]]^{-1/2}\exp[-\frac{v^2}{3ar^3t^4}[1+\frac{2}{rt}]^{-1}]\,. \quad (2\text{-}56)$$

From Eq. (2- the mean-squared displacement and the mean-squared velocity

$$<x^2(t)>=\frac{\alpha rt^4}{4}[1+\frac{4}{3rt}],\ <v^2(t)>=\frac{3ar^3t^4}{2}[1+\frac{2}{rt}]\,. \quad (2\text{-}57)$$

*2-5 Moment equation, kurtosis, correlation coefficient, and moment*

To this end we derive the moment equation for $\mu_{m,n}$ of distribution $P(x,v,t)$

$$\frac{d\mu_{m,n}}{dt}=-m\mu_{m-1,n+1}+mn\mu_{m-1,n}+rn\mu_{m,n}-em(n-1)\mu_{m-2,n}+\beta n\mu_{m+1,n-1}-mnac(t)\mu_{m-1,n-1}+n(n-1)ab(t)\mu_{m,n-2}\,. \quad (2\text{-}58)$$

Here $\mu_{m,n}=\int_{-\infty}^{+\infty}dx\int_{-\infty}^{+\infty}dvx^mv^nP(x,v,t)$. In order to get the accuracy of Gaussian distribution, we use its higher moments, $\mu_{2m,0}(t)=(2m-1)!![\mu_{2,0}(t)]^m$. The kurtosis for displacement and velocity are, respectively, given by

$$K_x=<x^4>/3<x^2>^2-1\quad,\quad K_v=<v^4>/3<v^2>^2-1. \quad (2\text{-}59)$$

We get the correlation coefficient as

$$\rho_{x,v}=<(x-<x>)\ (v-<v>)>/\sigma_x\sigma_v\,. \quad (2\text{-}60)$$

Here a passive particle is initially started at $x=x_0, v=v_0$. $\sigma_x$ ($\sigma_v$) is the root-mean-squared displacement (velocity) of joint probability density.

**Table 1.** Approximate values of the kurtosis, the correlation coefficient, and the moment $\mu_{2,2}$ in three-time domains, where a passive particle is initially started at $x=x_0, v=v_0$.

| time domains | variables | $K_x,\ K_v$ | $\rho_{x,v}$ | $\mu_{2,2}$ |
|---|---|---|---|---|
| $t<<\tau$ | $x$ | $\frac{\beta^2\tau^4x_0^4}{a^2}t^{-4}+\frac{\beta\tau^2x_0^2}{a}t^{-2}$ | $\frac{\beta^{1/2}\tau^{3/2}x_0v_0}{ar^{1/2}}t^{-3}$ | $\frac{a^2}{\tau(1+4et)}t^5$ |
| | $v$ | $\frac{\tau^2v_0^4}{a^2}t^{-8}+\frac{\tau v_0^2}{a}t^{-4}$ | | |
| $t >> \tau$ | $x$ | $\frac{\tau^2 x_0^4}{a^2 r^2} t^{-8} + + \frac{\tau x_0^2}{ar} t^{-4}$ | $\frac{\tau^{1/2} x_0 v_0}{ar} t^{-7/2}$ | $\frac{a^2}{\tau(1+4et)} t^5$ |
| | $v$ | $\frac{x_0^4}{a^2 r^4} t^{-8} + \frac{v_0^2}{ar^2} t^{-4}$ | | |
| $\tau = 0$ | $x$ | $\frac{x_0^4}{a^2 r^2} t^{-8} + \frac{x_0^2}{ar} t^{-4}$ | $\frac{x_0 v_0}{ar^2} t^{-4}$ | $\frac{a^2}{\tau(1+4et)} t^5$ |
| | $v$ | $\frac{v_0^4}{a^2 r^6} t^{-8} + \frac{v_0^2}{ar^3} t^{-4}$ | | |

**Table 2.** Approximate values of the kurtosis, the correlation coefficient, and the moment $\mu_{2,2}$ in three-time domains, where a passive particle with perturbative force $+cx^3$ is initially started at $x = x_0, v = v_0$ at time $t = 0$.

| time domains | variables | $K_x,\ K_v$ | $\rho_{x,v}$ | $\mu_{2,2}$ |
|---|---|---|---|---|
| $t << \tau$ | $x$ | $\frac{\tau^2 x_0^4}{a^2} t^{-8} + \frac{\tau x_0^2}{a} t^{-4}$ | $\frac{\tau x_0 v_0}{ar^{1/2}} t^{-4}$ | $\frac{a^2}{\tau(1-2rt)} t^5$ |
| | $v$ | $\frac{\tau^2 v_0^4}{a^2 r^2} t^{-8} + \frac{\tau v_0^2}{ar} t^{-4}$ | | |
| $t >> \tau$ | $x$ | $\frac{\tau^2 x_0^4}{a^2} t^{-8} + \frac{\tau x_0^2}{a} t^{-4}$ | $\frac{\tau^{1/2} x_0 v_0}{ar} t^{-7/2}$ | $\frac{a^2}{\tau(1-2rt)} t^6$ |
| | $v$ | $\frac{v_0^4}{a^2 r^4} t^{-6} + \frac{v_0^2}{ar^2} t^{-3}$ | | |
| $\tau = 0$ | $x$ | $\frac{x_0^4}{a^2 r^2} t^{-8} + \frac{x_0^2}{ar} t^{-4}$ | $\frac{x_0 v_0}{ar^2} t^{-4}$ | $\frac{a^2}{\tau(1-2rt)} t^6$ |
| | $v$ | $\frac{v_0^4}{a^2 r^6} t^{-8} + \frac{v_0^2}{ar^3} t^{-4}$ | | |

**Table 3.** Approximate values of the Kurtosis, the correlation coefficient, and the moment $\mu_{2,2}$ in three-time domains, where a passive particle with no harmonic and viscous forces $-\beta x - rv$ is initially started at $x = x_0, v = v_0$.

| time domains | variables | $K_x,\ K_v$ | $\rho_{x,v}$ | $\mu_{2,2}$ |
|---|---|---|---|---|
| $t << \tau$ | $x$ | $\frac{\tau^2 x_0^4}{a^2} t^{-8} + \frac{\tau x_0^2}{a} t^{-4}$ | $\frac{\tau x_0 v_0}{a} t^{-6}$ | $\frac{a^2}{\tau^2} t^6$ |
| | $v$ | $\frac{\tau^2 v_0^4}{a^2} t^{-4} + \frac{\tau v_0^2}{a} t^{-2}$ | | |
| $t >> \tau$ | $x$ | $\frac{x_0^4}{a^2} t^{-6} + \frac{x_0^4}{a} t^{-3}$ | $\frac{x_0 v_0}{a} t^{-4}$ | $\frac{a^2}{\tau^2} t^5$ |
| | $v$ | $\frac{v_0^4}{a^2} t^{-2} + \frac{v_0^2}{a} t^{-1}$ | | |
| $\tau = 0$ | $x$ | $\frac{x_0^4}{a^2} t^{-6} + \frac{x_0^4}{a} t^{-3}$ | $\frac{x_0 v_0}{a} t^{-4}$ | $\frac{a^2}{\tau^2} t^5$ |
| | $v$ | $\frac{v_0^4}{a^2} t^{-2} + \frac{v_0^2}{a} t^{-1}$ | | |

**Table 4.** Approximate values of the Kurtosis, the correlation coefficient, and the moment $\mu_{2,2}$ for a passive particle with (a) harmonic and viscous forces $-\beta x - rv$, (b) perturbative force $+cx^3$ (c) no harmonic and viscous forces $-\beta x - rv$, in three-time domains $t << \tau$ , $t >> \tau$ , and $\tau = 0$. Here V, MSD (MSV), and JPD is denoted the variable, mean squared displacement (velocity), and joint probability density, respectively.

| time domains | V | MSD MSV JPD | Force | | |
|---|---|---|---|---|---|
| | | | (a) harmonic and viscous forces | (b) perturbative force | (c) no harmonic and viscous forces |
| $t << \tau$ | $x$ | JPD | $\exp[-\frac{2\beta\tau x^2}{\alpha t^2}]$ | $\exp[-\frac{2\tau x^2}{at^4}]$ | $\exp[-\frac{4\tau x^2}{at^4}]$ |
| | | MSP | $\frac{\alpha}{4\beta\tau} t^2$ | $\frac{at^4}{4\tau}$ | $\frac{at^4}{8\tau}$ |
| | $v$ | JPD | $\exp[-\frac{2\tau v^2}{art^4}]$ | $\exp[-3\frac{\tau v^2}{2art^4}]$ | $\exp[-\frac{\tau v^2}{at^2}]$ |
| | | MSV | $\frac{art^4}{2\tau}$ | $\frac{art^4}{3\tau}$ | $\frac{at^2}{2\tau}$ |
| $t >> \tau$ | $x$ | JPD | $\exp[-\frac{3\tau x^2}{2at^4}]$ | $\exp[-\frac{3\tau v^2}{2at^4}]$ | $\exp[-\frac{3x^2}{2at^3}]$ |
| | | MSP | $\frac{at^4}{3\tau}$ | $\frac{at^4}{3\tau}$ | $\frac{at^3}{3}$ |
| | $v$ | JPD | $\exp[-\frac{x^2}{2ar^2t^3}]$ | $\exp[-\frac{v^2}{6ar^2t^3}]$ | $\exp[-\frac{v^2}{2at}]$ |
| | | MSV | $ar^2t^3$ | $3ar^2t^3$ | $at$ |
| $\tau = 0$ | $x$ | JPD | $\exp[-\frac{2x^2}{\alpha rt^4}]$ | $\exp[-\frac{2v^2}{art^4}]$ | $\exp[-\frac{3x^2}{at^3}]$ |
| | | MSD | $\frac{\alpha rt^4}{4}$ | $\frac{art^4}{4}$ | $\frac{at^3}{6}$ |
| | $v$ | JPD | $\exp[-\frac{v^2}{3ar^3t^4}]$ | $\exp[-\frac{v^2}{3ar^3t^4}]$ | $\exp[-\frac{v^2}{at}]$ |
| | | MSV | $\frac{3ar^3t^4}{2}$ | $\frac{3ar^3t^4}{2}$ | $\frac{at}{2}$ |

## 3. Navier-Stokes equation with harmonic force and magnetic field

### *3-1 Fokker-Planck equation in velocity and magnetic field*

In this section, we derive Fokker-Planck equation. The modified equations of motion for a conducting fluid in a magnetic field are expressed as follows [10, 11, 16]:

$$\frac{d}{dt}v(t) + v(t)\cdot\nabla v(t) + h(t)\cdot\nabla h(t) = e\nabla^2 v(t) - r_1 v(t) - \beta x(t) + g_v(t) + g_h(t) \qquad (3\text{-}1)$$

$$\frac{d}{dt}h(t) + h(t)\cdot\nabla v(t) - v(t)\cdot\nabla h(t) = \varepsilon\nabla^2 v(t) - r_2 h(t) - kx(t) + g_v(t) + g_h(t)\cdot \qquad (3\text{-}2)$$

We introduce that the incompressibility condition $\nabla\cdot v(t) = 0$, and we apply this condition to Eq. (3-1) and Eq. (3-2). Here, the parameter $e$ , $\varepsilon$ denote the drag coefficient, $-r_1 v(t)$ , $-r_2 h(t)$ the viscous force, and $-\beta x(t)$, $-kx(t)$ a harmonic force acting on the particle in velocity, magnetic field. In Eq. (3-1) and Eq. (3-2). We now add the random forces $g_v(t)$ and $g_h(t)$ that activates the motion of the particle by the fluctuation of the particle as follows:

$$< g_i(t) >= 0 \,, < g_i(t) g_j(t') >= g_i^2 g(t-t') \text{ for } i = v, h\,. \qquad (3\text{-}3)$$

Here $g(t-t') = \frac{1}{2\tau}\exp(-\frac{|t-t'|}{\tau})$, $g_v^2 = 2r_1 k_B T_v$ and $g_h^2 = 2r_2 k_B T_h$. Here $g_v$, $g_h$ denote the coupling strength, $r_1$, $r_2$ frictional constant, $T_v$ and $T_h$ temperature, $k_B$ Boltzmann constant, and $\tau$ characteristic time. The velocity and the magnetic field are governed two hydro-magnetic equations, Eq. (3-1) and Eq. (3-2), and we will later solve approximate solutions of these equations.

By taking time derivatives of the joint probability density, the joint probability density $P(x(t), v(t), h(t), t)$ for displacement $x(t)$ , velocity $v(t)$ , and magnetic field $h(t)$ is defined by $P(x(t), v(t), h(t), t) =< \delta(x - x(t))\delta(v - v(t))\delta(h - h(t)) >$. We assume that the passive particle is at rest at time $t = 0$ at initial state. Inserting Eq. (3-1) and Eq. (3-2) into $P(x(t), v(t), h(t), t)$, we write two equations for $P(x(t), v(t), h(t), t) \equiv P(x, v, h, t)$ and $\delta(x - x(t)) \equiv \delta_x$ as

$$\frac{\partial}{\partial t}P(x,v,h,t) = -\frac{\partial}{\partial x} < \frac{\partial x}{\partial t}\delta_x\delta_v\delta_h > -\frac{\partial}{\partial h} < g_h(t)\delta_x\delta_v\delta_h >$$
$$-\frac{\partial}{\partial v} < [-v\nabla v + v\nabla h + e\nabla^2 v - r_1 v - \beta x + g_v(t)]\delta_x\delta_v\delta_h > \qquad (3\text{-}4)$$

$$\frac{\partial}{\partial t}P(x,v,h,t) = -\frac{\partial}{\partial x} < \frac{\partial x}{\partial t}\delta_x\delta_v\delta_h > -\frac{\partial}{\partial v} < [-kx + g_v(t)]\delta_x\delta_v\delta_h >$$

$$-\frac{\partial}{\partial h}<[-h\nabla v+v\nabla h+\varepsilon\nabla^2 h-r_2 h+g_h(t)]\delta_x\delta_v\delta_h>\cdot \tag{3-5}$$

Finally, manipulating over integrals [24], Eq. (3-4) and Eq. (3-5) for $P(x,v,h,t)\equiv P$ are, respectively, derived like

$$\frac{\partial}{\partial t}P=-v\frac{\partial}{\partial x}P+\alpha_2 a(t)\frac{\partial^2}{\partial h^2}P+\frac{d}{dv}v\frac{d}{dx}vP-\frac{d}{dv}v\frac{d}{dx}hP-e\frac{d}{dv}\frac{d^2}{dx^2}vP+r_1\frac{d}{dv}vP$$

$$+\beta x\frac{\partial}{\partial v}P-\alpha_1 b(t)\frac{\partial^2}{\partial x\partial v}P+\alpha_1 a(t)\frac{\partial^2}{\partial v^2}P \tag{3-6}$$

$$\frac{\partial}{\partial t}P=-v\frac{\partial}{\partial x}P+kx\frac{\partial}{\partial v}P-a_1 d(t)\frac{\partial^2}{\partial x\partial v}P+a_1 c(t)\frac{\partial^2}{\partial v^2}P+\frac{d}{dh}h\frac{d}{dx}vP-\frac{d}{dh}v\frac{d}{dx}hP$$

$$-\varepsilon\frac{d}{dh}\frac{d^2}{dx^2}hP+r_2\frac{d}{dh}hP+a_2 c(t)\frac{\partial^2}{\partial h^2}P\cdot \tag{3-7}$$

Here $\alpha_1=a_1=g_v^2/2$ , $\alpha_2=a_2=g_h^2/2$ , $a(t)=c(t)=1-\exp(-t/\tau)$ and $b(t)=d(t)=(t+\tau)\quad\exp(-t/\tau)-\tau$ . The above equation is called the Fokker-Planck equation we derived, as mentioned in Introduction.

In order to find the joint probability density, we introduce the triple Fourier transform

$$P(\xi,\eta,\nu,t)=\int_{-\infty}^{+\infty}dx\int_{-\infty}^{+\infty}dv\int_{-\infty}^{+\infty}dh\exp(-i\xi x-i\eta v-i\nu h)P(x,v,h,t)\cdot \tag{3-8}$$

Taking Fourier transforms, the Fourier transforms of Eq. (3-6) and Eq. (3-7) are

$$\frac{\partial}{\partial t}P(\xi,\nu,\eta,t)=[-\beta\nu\frac{\partial}{\partial\xi}+[\xi-r\nu-e\xi^2]\frac{\partial}{\partial\nu}+\nu\xi\frac{\partial^2}{\partial\xi\partial\nu}-\xi\eta\frac{\partial^2}{\partial\nu\partial\eta}]P(\xi,\nu,\eta,t)$$

$$+[\alpha_1 b(t)\xi\eta-\alpha_1 a(t)\nu^2-\alpha_2 a(t)\eta^2]P(\xi,\nu,\eta,t)\,. \tag{3-9}$$

$$\frac{\partial}{\partial t}P(\xi,\nu,\eta,t)=[-k\nu\frac{\partial}{\partial\xi}+\xi\frac{\partial}{\partial\nu}-[\varepsilon\xi^2+r_2]\eta\frac{\partial}{\partial\eta}]P(\xi,\nu,\eta,t)$$

$$+[a_1 d(t)\xi\eta-a_1 c(t)\nu^2-a_2 c(t)\eta^2]P(\xi,\nu,\eta,t)\,. \tag{3-10}$$

*3-2 $P(x,t)$, $P(v,t)$ and $P(h,t)$ in short-time domain*

In this section, we will find $P(x,t)$, $P(v,t)$ and $P(h,t)$ in short-time domain $t<<\tau$ .

*3-2-1 $P(x,t)$, $P(v,t)$ and $P(h,t)$ of velocity*. In order to obtain the special solutions for $\xi$, $\nu$, $\eta$ by variable separation from Eq. (3-9), three equations of velocity are as follows:

$$\frac{\partial}{\partial t}P(\xi,t)=-\beta\nu\frac{\partial}{\partial\nu}P(\xi,t)+\frac{1}{3}[\alpha_1 b(t)\xi\nu-\alpha_1 a(t)\nu^2-\alpha_2 a(t)\eta^2]P(\xi,t)+fP(\xi,t) \tag{3-11}$$

$$\frac{\partial}{\partial t}P(\nu,t)=(\xi-r_1\nu-e\xi^2\nu+\xi\nu D_\nu)\frac{\partial}{\partial\nu}P(\nu,t)+\frac{1}{3}[\alpha_1 b(t)\xi\nu-\alpha_1 a(t)\nu^2-\alpha_2 a(t)\eta^2]P(\nu,t)-\frac{f}{2}P(\nu,t) \tag{3-12}$$

$$\frac{\partial}{\partial t}P(\eta,t)=-\xi\eta D_\nu\frac{\partial}{\partial\eta}P(\eta,t)+\frac{1}{3}[\alpha_1 b(t)\xi\nu-\alpha_1 a(t)\nu^2-\alpha_2 a(t)\eta^2]P(\eta,t)-\frac{f}{2}P(\eta,t) \tag{3-13}$$

Here $D_\nu\equiv\partial/\partial\nu$ and $f$ denotes the separation constant. In steady state, introducing $\frac{\partial}{\partial t}P(\xi,t)=0$ and $P(\xi,t)\to P^{st}(\xi,t)$, we get Eq. (3-11) as

$$P^{st}(\xi,t)=\exp[\frac{1}{3\beta\nu}[\alpha_1 b(t)\nu\frac{\xi^2}{2}-\alpha_1 a(t)\nu^2\xi-\alpha_2 a(t)\eta^2\xi]P(\xi,t)+\frac{f}{2\beta\nu}\xi P(\xi,t)] \tag{3-14}$$

In order to find the solutions of joint functions for $\xi$ from $P(\xi,t)\equiv Q(\xi,t)P^{st}(\xi,t)$, we obtain the distribution functions via the calculation including terms up to order $t^2/\tau^2$, that is,

$$P(\xi,t)=Q(\xi,t)P^{st}(\xi,t) \tag{3-15}$$

$$Q(\xi,t)=R(\xi,t)Q^{st}(\xi,t)=R(\xi,t)\exp[-\frac{1}{3(\beta\nu)^2}[\alpha_1 b'(t)\nu\frac{\xi^3}{6}-a'(t)[\alpha_1\nu^2-\alpha_2\eta^2]\frac{\xi^2}{2}] \tag{3-16}$$

$$R(\xi,t)=S(\xi,t)R^{st}(\xi,t)=S(\xi,t)\exp[\frac{1}{3(\beta\nu)^3}[\alpha_1 b''(t)\nu\frac{\xi^4}{24}-a''(t)[\alpha_1\nu^2-\alpha_2\eta^2]\frac{\xi^3}{6}] \qquad (3\text{-}17)$$

$$S(\xi,t)=T(\xi,t)S^{st}(\xi,t)=T(\xi,t)\exp[-\frac{1}{3(\beta\nu)^4}[\alpha_1 b'''(t)\nu\frac{\xi^5}{120}]]\cdot \qquad (3\text{-}18)$$

Here $a'(t)\equiv d/dt$ and $a''(t)\equiv d^2/dt^2$. Discarding terms proportional to $1/\tau^3$ and taking the solutions as arbitrary functions of variable $t-\xi/\beta\nu$, the arbitrary function $T(\xi,t)$ becomes $T(\xi,t)=\Theta[t-\xi/\ \beta\nu]$. Consequently, we find that

$$P(\xi,t)=\Theta[t-\xi/\beta\nu]S^{st}(\xi,t)R^{st}(\xi,t)Q^{st}(\xi,t)P^{st}(\xi,t) \qquad (3\text{-}19)$$

By the similar method (Eq. (3-14) – (3-19)) from Eq. (3-12) and Eq. (3-13), we get Fourier transforms of probability density for $\nu,\ \eta$ as

$$P(\nu,t)=\Theta[t+\nu/(\xi-r_1\nu-e\xi^2\nu+\xi\nu D_\nu)]S^{st}(\nu,t)R^{st}(\nu,t)Q^{st}(\nu,t)P^{st}(\nu,t) \qquad (3\text{-}20)$$

$$P(\eta,t)=\Theta[t-\eta/\xi\nu D_\eta]S^{st}(\eta,t)R^{st}(\eta,t)Q^{st}(\eta,t)P^{st}(\eta,t)\cdot \qquad (3\text{-}21)$$

Therefore, from Eq. (3-19), Eq. (3-20), and Eq. (3-21), we calculate that

$$P(\xi,\nu,\eta,t)=P(\xi,t)P(\nu,t)P(\eta,t)$$

$$=\exp[-\frac{\alpha_1 t^4}{4\tau}[1-\frac{\tau}{t}]\xi^2-\frac{\alpha_1 t^3}{6\tau}[1+\frac{2\beta}{t}]\nu^2-\frac{\alpha_2 t^3}{2\tau}[1-\frac{\tau}{t}]\eta^2]\,. \qquad (3\text{-}22)$$

Using the inverse Fourier transform, we have

$$P(x,t)=[\pi\frac{\alpha_1 t^4}{\tau}[1-\frac{\tau}{t}]]^{-1/2}\exp[-\frac{\tau x^2}{\alpha_1 t^4}[1-\frac{\tau}{t}]^{-1}] \qquad (3\text{-}23)$$

$$P(\nu,t)=[2\pi\frac{\alpha_1 t^3}{3\tau}[1+\frac{2\beta}{t}]]^{-1/2}\exp[-\frac{3\tau\nu^2}{2\alpha_1 t^3}[1+\frac{2\beta}{t}]^{-1}] \qquad (3\text{-}24)$$

$$P(h,t)=[2\pi\frac{\alpha_2 t^3}{\tau}[1-\frac{\tau}{t}]]^{-1/2}\exp[-\frac{\tau h^2}{2\alpha_2 t^3}[1-\frac{\tau}{t}]^{-1}]\,. \qquad (3\text{-}25)$$

The mean squared displacement for $P(x,t)$ are, respectively, given by

$$<x^2>=\frac{\alpha_1 t^4}{2\tau}[1-\frac{\tau}{t}]\ ,\ <\nu^2>=\frac{\alpha_1 t^3}{3\tau}[1+\frac{2\beta}{t}]\ ,\ <h^2>=\frac{\alpha_2 t^3}{\tau}[1-\frac{\tau}{t}]\,. \qquad (3\text{-}26)$$

*3-2-2 $P(x,t)$, $P(v,t)$ and $P(h,t)$ of magnetic field*. From Eq. (3-10), in order to obtain the special solutions for $\xi,\ \nu,\ \eta$ by variable separation three equations of magnetic field are as follows:

$$\frac{\partial}{\partial t}P(\xi,t)=-k\nu\frac{\partial}{\partial\xi}P(\xi,t)+\frac{1}{3}[a_1 d(t)\xi\nu-a_1 c(t)\nu^2-a_2 c(t)\eta^2]P(\xi,t)+hP(\xi,t) \qquad (3\text{-}27)$$

$$\frac{\partial}{\partial t}P(\nu,t)=\xi\frac{\partial}{\partial\nu}P(\nu,t)+\frac{1}{3}[a_1 d(t)\xi\nu-a_1 c(t)\nu^2-a_2 c(t)\eta^2]P(\nu,t)-\frac{h}{2}P(\nu,t) \qquad (3\text{-}28)$$

$$\frac{\partial}{\partial t}P(\eta,t)=-(\varepsilon\xi^2\eta+r_2\eta)\frac{\partial}{\partial\eta}P(\eta,t)+\frac{1}{3}[a_1 d(t)\xi\nu-a_1 c(t)\nu^2-a_2 c(t)\eta^2]P(\eta,t)-\frac{h}{2}P(\eta,t)\cdot \qquad (3\text{-}29)$$

In steady state, introducing $\frac{\partial}{\partial t}P(\nu,t)=0$ and $P(\nu,t)\to P^{st}(\nu,t)$, we get Eq. (3-28) as

$$P^{st}(\nu,t)=\exp[\frac{1}{3\xi}[-a_1 d(t)\xi\frac{\nu^2}{2}+a_1 c(t)\frac{\nu^3}{3}+a_2 c(t)\eta^2\nu]P(\nu,t)-\frac{h}{2\xi}\nu P(\nu,t)] \qquad (3\text{-}30)$$

In order to find the solutions of joint functions for $\nu$ from $P(\nu,t)\equiv Q(\nu,t)P^{st}(\nu,t)$, we obtain the distribution functions via the calculation including terms up to order $t^2/\tau^2$, that is,

$$P(\nu,t)=Q(\nu,t)P^{st}(\nu,t) \qquad (3\text{-}31)$$

$$Q(\nu,t)=R(\nu,t)Q^{st}(\nu,t)=R(\nu,t)\exp[\frac{1}{3\xi^2}[-a_1 d'(t)\xi\frac{\nu^3}{6}+a_1 c'(t)\frac{\nu^4}{12}+a_2 c'(t)\eta^2\frac{\nu^2}{2}]] \qquad (3\text{-}32)$$

$$R(\nu,t)=S(\nu,t)R^{st}(\nu,t)=S(\nu,t)\exp[\frac{1}{3\xi^3}[-a_1d''(t)\xi\frac{\nu^4}{24}+a_1c''(t)\frac{\nu^5}{60}+a_2c''(t)\eta^2\frac{\nu^3}{6}]] \qquad (3\text{-}33)$$

$$S(\nu,t)=T(\nu,t)S^{st}(\nu,t)=T(\nu,t)\exp[-\frac{1}{3\xi^3}a_1d'''(t)\frac{\nu^4}{24}]\cdot \qquad (3\text{-}34)$$

Discarding terms proportional to $1/\tau^3$ and taking the solutions as arbitrary functions of variable $t+\nu/\xi$, the arbitrary function $T(\xi,t)$ becomes $T(\nu,t)=\Theta[t+\nu/\xi]$. Consequently, we find that

$$P(\nu,t)=\Theta[t+\nu/\xi]S^{st}(\nu,t)R^{st}(\nu,t)Q^{st}(\nu,t)P^{st}(\nu,t) \qquad (3\text{-}35)$$

By the similar method (Eq. (3-30) – (3-35)) from Eq. (3-27) and Eq. (3-29), we get Fourier transforms of probability density for $\xi,\ \eta$ as

$$P(\xi,t)=\Theta[t-\xi/\varepsilon\nu]S^{st}(\xi,t)R^{st}(\xi,t)Q^{st}(\xi,t)P^{st}(\xi,t) \qquad (3\text{-}36)$$

$$P(\eta,t)=\Theta[t-\ln\eta/(\lambda\xi^2+r_2)]S^{st}(\eta,t)R^{st}(\eta,t)Q^{st}(\eta,t)P^{st}(\eta,t)\cdot \qquad (3\text{-}37)$$

Therefore, from Eq. (3-35), Eq. (3-36), and Eq. (3-37), we calculate that

$$P(\xi,\nu,\eta,t)=P(\xi,t)P(\nu,t)P(\eta,t)$$

$$=\exp[-\frac{a_1t^4}{24\tau}[1-\frac{4t}{15\tau}]\xi^2-\frac{a_1t}{6\tau}[1-\frac{t}{3\tau}]\nu^2-\frac{a_2t^2}{2\tau}[1-\frac{t}{a_2\tau}]\eta^2]. \qquad (3\text{-}38)$$

Using the inverse Fourier transform, we have

$$P(x,t)=[\pi\frac{a_1t^4}{6\tau}[1-\frac{4t}{15\tau}]]^{-1/2}\exp[-\frac{6\tau x^2}{a_1t^4}[1-\frac{4t}{15\tau}]^{-1}] \qquad (3\text{-}39)$$

$$P(v,t)=[2\pi\frac{a_1t}{3\tau}[1-\frac{t}{3\tau}]]^{-1/2}\exp[-\frac{3\tau v^2}{2a_1t}[1-\frac{t}{3\tau}]^{-1}] \qquad (3\text{-}40)$$

$$P(h,t)=[2\pi\frac{a_2t^2}{\tau}[1-\frac{t}{a_2\tau}]]^{-1/2}\exp[-\frac{\tau h^2}{2a_2t^3}[1-\frac{t}{a_2\tau}]^{-1}]. \qquad (3\text{-}41)$$

The mean squared displacement, velocity, and magnetic field for $P(v,t)$, $P(v,t)$ and $P(h,t)$ are, respectively, given by

$$<x^2>=\frac{a_1t^4}{12\tau}[1-\frac{4t}{15\tau}],\ <v^2>=\frac{a_1t}{3\tau}[1-\frac{t}{3\tau}],\ <h^2>=\frac{a_2t^2}{\tau}[1-\frac{t}{a_2\tau}]. \qquad (3\text{-}42)$$

*3-3 $P(x,t)$, $P(v,t)$, and $P(h,t)$ in long-time domain*

In this section, we will find $P(x,t)$, $P(v,t)$ and $P(h,t)$ in long-time domain $t>>\tau$.

*3.3.1 $P(x,t)$, $P(v,t)$ and $P(h,t)$ of velocity.* In long-time domain, we write approximate equation for $\xi,\ \nu,\ \eta$ from Eq. (3-11)-(3-13) as

$$\frac{\partial}{\partial t}P_\xi(\xi,t)\cong\frac{1}{3}[\alpha_1b(t)\xi\nu-\alpha_1a(t)\nu^2-\alpha_2a(t)\eta^2]P_\xi(\xi,t) \qquad (3\text{-}44)$$

$$\frac{\partial}{\partial t}P_\nu(\nu,t)\cong\frac{1}{3}[\alpha_1b(t)\xi\nu-\alpha_1a(t)\nu^2-\alpha_2a(t)\eta^2]P_\nu(\nu,t) \qquad (3\text{-}45)$$

$$\frac{\partial}{\partial t}P_\eta(\eta,t)\cong\frac{1}{3}[\alpha_1b(t)\xi\nu-\alpha_1a(t)\nu^2-\alpha_2a(t)\eta^2]P_\eta(\eta,t)\cdot \qquad (3\text{-}46)$$

In steady state, as $\frac{\partial}{\partial t}P(\xi,t)=0$, $\frac{\partial}{\partial t}P(\nu,t)=0$ and $\frac{\partial}{\partial t}P(\eta,t)=0$, the probability density $P(\xi,t)$, $P(\nu,t)$ and $P(\eta,t)$ become $P^{st}(\xi,t)$, $P^{st}(\nu,t)$ and $P^{st}(\eta,t)$. Then we calculate $P_\xi^{st}(\xi,t)$, $P_\eta^{st}(\nu,t)$ and $P_\eta^{st}(\eta,t)$ as

$$P_\xi^{st}(\xi,t)=\exp[\frac{1}{3}\int[\alpha_1b(t)\xi\nu-\alpha_1a(t)\nu^2-\alpha_2a(t)\eta^2]\,dt] \qquad (3\text{-}47)$$

$$P_{\nu}^{st}(\nu,t)=\exp[\frac{1}{3}\int[\alpha_1 b(t)\xi\nu-\alpha_1 a(t)\nu^2-\alpha_2 a(t)\eta^2]\ dt] \quad (3\text{-}48)$$

$$P_{\eta}^{st}(\eta,t)=\exp[\frac{1}{3}\int[\alpha_1 b(t)\xi\nu-\alpha_1 a(t)\nu^2-\alpha_2 a(t)\eta^2]\ dt]\cdot \quad (3\text{-}49)$$

Here $\int a(t)dt=t-\tau$ , $a(t)=1$ and $\int b(t)dt=-\tau t$ , $b(t)=-\tau$ in long-time domain. We find $Q_{\xi}^{st}(\xi,t)$ , $Q_{\nu}^{st}(\nu,t)$ and $Q_{\eta}^{st}(\eta,t)$ for $\xi,\ \nu,\ \eta$, from $P_{\xi}(\xi,t)\equiv Q_{\xi}(\xi,t)P_{\xi}^{st}(\xi,t)$, that is,

$$Q_{\xi}^{st}(\xi,t)=\exp[\frac{1}{3}\int[-\alpha_1 b(t)\xi\nu+\alpha_1 a(t)\nu^2+\alpha_2 a(t)\eta^2]\ dt] \quad (3\text{-}50)$$

$$Q_{\nu}^{st}(\nu,t)=\exp[\frac{1}{3}\int[-\alpha_1 b(t)\xi\nu+\alpha_1 a(t)\nu^2+\alpha_2 a(t)\eta^2]\ dt] \quad (3\text{-}51)$$

$$Q_{\eta}^{st}(\eta,t)=\exp[\frac{1}{3}\int[-\alpha_1 b(t)\xi\nu+\alpha_1 a(t)\nu^2+\alpha_2 a(t)\eta^2]\ dt]\ \cdot \quad (3\text{-}52)$$

From Eq. (3-11) for $\xi$ in long-time domain, Fourier transform of the probability density in steady state is calculated as $P^{st}(\xi,t)=\exp[\frac{1}{3\beta\nu}[\alpha_1 b(t)\nu\frac{\xi^2}{2}-a(t)[\alpha_1\nu^2-\alpha_2\eta^2]\xi]+\frac{D_\nu}{3(\beta\nu)^2}\ [\alpha_1 b(t)\eta\frac{\xi^3}{3}-a(t)[\alpha_1\nu^2-\alpha_2\eta^2]\frac{\xi^2}{2}]$. It is clear in long-time domain that Fourier transform of probability density $P^{st}(\xi,t)$ for $\xi$ and Eq. (3-14) should be the same. The probability density $P(\xi,t)$ is given by

$$P(\xi,t)=\Theta[t-\xi/\beta\nu]Q_{\xi}^{st}(\xi,t)P^{st}(\xi,t)\,. \quad (3\text{-}53)$$

Later, we substitute the calculated values of Eq. (3-14) and Eq. (3-50) into Eq. (3-53), and then we get $P(\xi,t)$ of Eq. (3-53). Using the similar method of $P(\xi,t)$ derived, we also get Fourier transforms of probability density for $\nu,\ \eta$ as

$$P(\nu,t)=\Theta[t+\nu/(\xi-r_1\nu-e\xi^2\nu+\xi\nu D_\nu)]Q_{\nu}^{st}(\nu,t)P^{st}(\nu,t) \quad (3\text{-}54)$$

$$P(\eta,t)=\Theta[t-\eta/\xi\nu D_\eta]Q_{\eta}^{st}(\eta,t)P^{st}(\eta,t)\,. \quad (3\text{-}55)$$

For long-time domain $t>>\tau$ , from Eq. (3-53), (3-54) and Eq. (3-55), we have

$$P(\xi,\nu,\eta,t)=P(\xi,t)P(\nu,t)P(\eta,t)=\exp[-\frac{2\alpha_1 t}{3}[1+\tau]\xi^2-\frac{2\alpha_1\beta^2t^3}{9}[1-\frac{3\tau}{\beta^2t}]\nu^2-\alpha_2 t\eta^2]\,. \quad (3\text{-}56)$$

Using the inverse Fourier transform, we have

$$P(x,t)=[2\pi\frac{4\alpha_1 t}{3}[1+\tau]]^{-1/2}\exp[-\frac{3x^2}{8\alpha_1 t}[1+\tau]^{-1}] \quad (3\text{-}57)$$

$$P(v,t)=[2\pi\frac{4\alpha_1\beta^2t^3}{9}[1-\frac{3\tau}{\beta^2t}]]^{-1/2}\exp[-\frac{9v^2}{8\alpha_1\beta^2t^3}[1-\frac{3\tau}{\beta^2t}]^{-1}] \quad (3\text{-}58)$$

$$P(h,t)=[4\pi\alpha_2 t]^{-1/2}\exp[-\frac{h^2}{4\alpha_2 t}]\,. \quad (3\text{-}59)$$

Mean squared displacement, velocity, and magnetic field for $P(v,t)$ , $P(v,t)$ and $P(h,t)$ are, respectively, given by

$$<x^2>=\frac{4\alpha_1 t}{3}[1+\tau]\,,\ <v^2>=\frac{4\alpha_1\beta^2t^3}{9}[1-\frac{3\tau}{\beta^2t}]\,,\ <h^2>=2\alpha_2 t\,. \quad (3\text{-}60)$$

*3.3.2* $P(x,t)$, $P(v,t)$, *and* $P(h,t)$ *of magnetic field*. In long-time domain, we write approximate equation for $\xi,\ \nu,\ \eta$ from Eq. (3-27)-(3-29) as

$$\frac{\partial}{\partial t}P_{\xi}(\xi,t)\cong\frac{1}{3}[a_1 d(t)\xi\nu-a_1 c(t)\nu^2-a_2 c(t)\eta^2]P_{\xi}(\xi,t) \quad (3\text{-}61)$$

$$\frac{\partial}{\partial t}P_{\nu}(\nu,t)\cong\frac{1}{3}[a_1 d(t)\xi\nu-a_1 c(t)\nu^2-a_2 c(t)\eta^2]P_{\nu}(\nu,t) \quad (3\text{-}62)$$

$$\frac{\partial}{\partial t}P_\eta(\eta,t) \cong \frac{1}{3}[a_1 d(t)\xi\nu - a_1 c(t)\nu^2 - a_2 c(t)\eta^2]P_\eta(\eta,t)\cdot \quad (3\text{-}63)$$

In steady state, introducing $\frac{\partial}{\partial t}P_\xi(\xi,t)=0$ for $\xi$ from Eq. (3-61) and $P_\xi(\xi,t)\to P_\xi^{st}(\xi,t)$, then we can calculate $P_\xi^{st}(\xi,t)$, $P_\eta^{st}(\nu,t)$ and $P^{st}(\eta,t)$ as

$$P_\xi^{st}(\xi,t)=\exp[\frac{1}{3}\int[a_1 d(t)\xi\nu - a_1 c(t)\nu^2 - a_2 c(t)\eta^2]\,dt] \quad (3\text{-}64)$$

$$P_\nu^{st}(\nu,t)=\exp[\frac{1}{3}\int[a_1 d(t)\xi\nu - a_1 c(t)\nu^2 - a_2 c(t)\eta^2]\,dt] \quad (3\text{-}65)$$

$$P_\eta^{st}(\eta,t)=\exp[\frac{1}{3}\int[a_1 d(t)\xi\nu - a_1 c(t)\nu^2 - a_2 c(t)\eta^2]\,dt]\cdot \quad (3\text{-}67)$$

We find $Q_\xi^{st}(\xi,t)$, $Q_\nu^{st}(\nu,t)$ and $Q_\eta^{st}(\eta,t)$ for $\xi,\ \nu,\ \eta$ from $P_\xi(\xi,t)\equiv Q_\xi(\xi,t)P_\xi^{st}(\xi,t)$, that is,

$$Q_\xi^{st}(\xi,t)=\exp[\frac{1}{3}\int[-a_1 d(t)\xi\nu + a_1 c(t)\nu^2 + a_2 c(t)\eta^2]\,dt] \quad (3\text{-}68)$$

$$Q_\nu^{st}(\nu,t)=\exp[\frac{1}{3}\int[-a_1 d(t)\xi\nu + a_1 c(t)\nu^2 + a_2 c(t)\eta^2]\,dt] \quad (3\text{-}69)$$

$$Q_\eta^{st}(\eta,t)=\exp[\frac{1}{3}\int[-a_1 b(t)\xi\nu + a_1 a(t)\nu^2 + a_2 a(t)\eta^2]\,dt]\cdot \quad (3\text{-}70)$$

From Eq. (3-28) for the magnetic field, Fourier transform of the probability density for $\nu$ in steady state is $P^{st}(\nu,t)=\exp[\frac{1}{3\xi}\ [-a_1 d(t)\xi\frac{\nu^2}{2}+a_1 c(t)\frac{\nu^3}{3}+a_2 c(t)\eta^2\nu]P(\nu,t)-\frac{h}{2\xi}\nu P(\nu,t)]$. It is also clear in long-time domain that Fourier transform of probability density $P^{st}(\nu,t)$ for $\nu$ and Eq. (3-30) should be the same. From Eq. (3-69) and Eq. (3-30), $P(\nu,t)$ is given by

$$P(\nu,t)=\Theta[t+\nu/\xi]Q_\nu^{st}(\nu,t)P^{st}(\nu,t)\,. \quad (3\text{-}71)$$

By the similar method, we get Fourier transforms of probability density for $\xi,\ \eta$ as

$$P(\xi,t)=\Theta[t-\xi/\varepsilon\nu]Q_\xi^{st}(\varsigma,t)P^{st}(\xi,t) \quad (3\text{-}72)$$

$$P(\eta,t)=\Theta[t-\ln\eta/(\lambda\xi^2+r_2)]Q_\eta^{st}(\eta,t)P^{st}(\eta,t)\,. \quad (3\text{-}73)$$

For long-time domain $t>>\tau$, from Eq. (3-71), (3-72) and Eq. (3-73), we have

$$P(\xi,\nu,\eta,t)=P(\xi,t)P(\nu,t)P(\eta,t)=\exp[-\frac{2a_2\tau t^2}{3}[1+\frac{a_1}{2a_2 t}]\xi^2-\frac{2a_1\varepsilon^2 t^3}{9}[1-\frac{3\tau}{\varepsilon^2 t}]\nu^2-\frac{a_2 t}{3}\eta^2]\,. \quad (3\text{-}74)$$

Using the inverse Fourier transform, we have

$$P(x,t)=[2\pi\frac{4a_2\tau t^2}{3}[1+\frac{a_1}{2a_2 t}]]^{-1/2}\exp[-\frac{3x^2}{8a_2\tau t^2}[1+\frac{a_1}{2a_2 t}]^{-1}] \quad (3\text{-}75)$$

$$P(v,t)=[2\pi\frac{4a_1\varepsilon^2 t^3}{9}[1-\frac{3\tau}{\varepsilon^2 t}]]^{-1/2}\exp[-\frac{9v^2}{8a_1\varepsilon^2 t^3}[1-\frac{3\tau}{\varepsilon^2 t}]^{-1}] \quad (3\text{-}76)$$

$$P(h,t)=[2\pi\frac{2a_2 t}{3}]^{-1/2}\exp[-\frac{3h^2}{4a_2 t}]\,. \quad (3\text{-}77)$$

Mean squared displacement, velocity, and magnetic field for $P(v,t)$, $P(v,t)$ and $P(h,t)$ are, respectively, given by

$$< x^2 >=\frac{4a_2\tau t^2}{3}[1+\frac{a_1}{2a_2 t}],\ < v^2 >=\frac{4a_1\varepsilon^2 t^3}{9}[1-\frac{3\tau}{\varepsilon^2 t}],\ < h^2 >=\frac{2a_2 t}{3}\,. \quad (3\text{-}78)$$

*3-4 $P(x,t)$, $P(v,t)$ and $P(h,t)$ in $\tau=0$*

In this section, we will find $P(x,t)$, $P(v,t)$, and $P(h,t)$ in time domain $\tau = 0$ ($t \to \infty$).

*3.4.1 $P(x,t)$, $P(v,t)$ and $P(h,t)$ of velocity*. In $\tau = 0$ domain ($a(t)=1, b(t)=0$), we write approximate equation from Eq. (3-11), Eq. (3-12) and Eq. (3-13) for $\xi,\ \nu,\ \eta$

$$\frac{\partial}{\partial t}P(\xi,t) = -\beta\nu\frac{\partial}{\partial\xi}P(\xi,t) - \frac{1}{3}[\alpha_1\nu^2 + \alpha_2\eta^2]P(\xi,t) \tag{3-79}$$

$$\frac{\partial}{\partial t}P(\nu,t) = (\xi - r_1\nu - e\xi^2\nu + \xi\nu D_\nu)\frac{\partial}{\partial\nu}P(\nu,t) - \frac{1}{3}[\alpha_1\nu^2 + \alpha_2\eta^2]P(\xi,t) \tag{3-80}$$

$$\frac{\partial}{\partial t}P(\eta,t) = -\xi\eta D_\nu\frac{\partial}{\partial\eta}P(\eta,t) - \frac{1}{3}[\alpha_1\nu^2 + \alpha_2\eta^2]P(\xi,t)\cdot \tag{3-81}$$

In steady state, we calculate $P^{st}(\xi,t)$, $P^{st}(\nu,t)$ and $P^{st}(\eta,t)$ as

$$P^{st}(\xi,t) = \exp[-\int\frac{1}{3\beta\nu}[\alpha_1\nu^2 + \alpha_2\eta^2]d\xi] \tag{3-82}$$

$$P^{st}(\nu,t) = \exp[-\int\frac{1}{3(\xi - r_1\nu - e\xi^2\nu + \xi\nu D_\nu)}[\alpha_1\nu^2 + \alpha_2\eta^2]d\nu] \tag{3-83}$$

$$P^{st}(\eta,t) = \exp[-\int\frac{1}{3\xi\eta D_\nu}[\alpha_1\nu^2 + \alpha_2\eta^2]d\eta]\cdot \tag{3-84}$$

We can find $P(\xi,t)$, $P(\nu,t)$ and $P(\eta,t)$ as

$$P(\xi,t) = \Theta_\xi[t - \xi/\beta\nu]P^{st}(\xi,t) \tag{3-85}$$

$$P(\nu,t) = \Theta_\xi[t + \nu/(\xi - r_1\nu - e\xi^2\nu + \xi\nu D_\nu)]P^{st}(\nu,t) \tag{3-86}$$

$$P(\eta,t) = \Theta[t - \eta/\xi\eta D_\nu)]P^{st}(\eta,t)\cdot \tag{3-87}$$

Thus we can calculate $P(\xi,\eta,t)$ from Eq. (3-85), (3-86) and (3-87) as

$$P(\xi,\eta,t) = P(\xi,t)P(\eta,t) = \exp[-\frac{\alpha_1 r_1 t^4}{8}[1+\frac{4}{3r_1t}]\xi^2 - \frac{\alpha_1 t}{2}[1+\frac{1}{4r_1t}]\nu^2 - \frac{\alpha_2 t^2}{6}\eta^2]\cdot \tag{3-88}$$

Using the inverse Fourier transform, $P(x,t)$, $P(v,t)$ and $P(h,t)$ are, respectively, presented by

$$P(x,t) = [\pi\frac{\alpha_1 r_1 t^4}{2}[1+\frac{4}{3r_1t}]]^{-1/2}\exp[-\frac{2x^2}{\alpha_1 r_1 t^4}[1+\frac{4}{3r_1t}]^{-1}] \tag{3-89}$$

$$P(v,t) = [2\pi\alpha_1 t[1+\frac{1}{4r_1t}]]^{-1/2}\exp[-\frac{v^2}{2\alpha_1 t}[1+\frac{1}{4r_1t}]^{-1}] \tag{3-90}$$

$$P(h,t) = [2\pi\frac{\alpha_2 t^2}{3}]^{-1/2}\exp[-\frac{3h^2}{2\alpha_2 t^2}]\cdot \tag{3-91}$$

Thus the mean-squared deviations are, respectively, given by

$$<x^2(t)> = \frac{\alpha_1 r_1 t^4}{4}[1+\frac{4}{3r_1t}],\ <v^2(t)> = \alpha_1 t[1+\frac{1}{4r_1t}],\ <\eta^2(t)> = \frac{\alpha_2 t^2}{3}\cdot \tag{3-92}$$

*3.4.2 $P(x,t)$, $P(v,t)$ and $P(h,t)$ of magnetic field*. In $\tau = 0$ domain ($c(t)=1, d(t)=0$), we write approximate equation from Eq. (3-27), (3-28) and (3-29) for $\xi,\ \nu,\ \eta$

$$\frac{\partial}{\partial t}P(\xi,t) = -k\nu\frac{\partial}{\partial\xi}P(\xi,t) - \frac{1}{3}[a_1\nu^2 + a_2\eta^2]P(\xi,t) \tag{3-93}$$

$$\frac{\partial}{\partial t}P(\nu,t) = \xi\frac{\partial}{\partial\nu}P(\nu,t) - \frac{1}{3}[a_1\nu^2 + a_2\eta^2]P(\nu,t) \tag{3-94}$$

$$\frac{\partial}{\partial t}P(\eta,t) = -(\varepsilon\xi^2\eta + r_2\eta)\frac{\partial}{\partial\eta}P(\eta,t) - \frac{1}{3}[a_1\nu^2 + a_2\eta^2]P(\eta,t)\cdot \tag{3-95}$$

In steady state, we calculate $P^{st}(\xi,t)$, $P^{st}(\nu,t)$ and $P^{st}(\eta,t)$ as

$$P^{st}(\xi,t)=\exp[-\int\frac{1}{3k\nu}[a_1\nu^2+a_2\eta^2]d\xi] \quad (3\text{-}96)$$

$$P^{st}(\nu,t)=\exp[\int\frac{1}{3\xi}[a_1\nu^2+a_2\eta^2]d\nu] \quad (3\text{-}97)$$

$$P^{st}(\eta,t)=\exp[-\int\frac{1}{3(\varepsilon\xi^2+r_2)\eta}[a_1\nu^2+a_2\eta^2]d\eta]\cdot \quad (3\text{-}98)$$

We can find $P(\xi,t), P(\nu,t)$ and $P(\eta,t)$ as

$$P(\xi,t)=\Theta[t-\xi/k\nu]P^{st}(\xi,t),\ P(\nu,t)=\Theta[t+\nu/\xi]P^{st}(\nu,t),\ P(\eta,t)=\Theta[t-\ln\eta/(\varepsilon\xi^2+r_2)]P^{st}(\eta,t). \quad (3\text{-}99)$$

Therefore, we can calculate $P(\xi,\eta,t)$ from Eq. (3-99) as

$$P(\xi,\eta,t)=P(\xi,t)P(\eta,t)=\exp[-\frac{a_1t^3}{18}\xi^2-\frac{a_1t}{2}\nu^2-\frac{a_2t}{3}\eta^2]\cdot \quad (3\text{-}100)$$

Using the inverse Fourier transform, $P(x,t)$, $P(v,t)$ and $P(h,t)$ are, respectively, presented by

$$P(x,t)=[2\pi\frac{a_1t^3}{9}]^{-1/2}\exp[-\frac{9x^2}{2a_1t^3}],\ P(v,t)=[2\pi a_1t]^{-1/2}\exp[-\frac{v^2}{2a_1t}],\ P(h,t)=[2\pi\frac{2a_2t}{3}]^{-1/2}\exp[-\frac{3h^2}{4a_2t}]\cdot \quad (3\text{-}101)$$

From Eq. (3-103), Eq. (3-104) and Eq. (3-105), we have the mean-squared deviations like

$$<x^2(t)>=\frac{a_1t^3}{9},\ <v^2(t)>=a_1t,\ <\eta^2(t)>=\frac{2a_2t}{3}. \quad (3\text{-}102)$$

*3-5. Moment equation, kurtosis, correlation coefficient, and moment*

To this end we derive the moment equation for $\mu^v_{l,m,n}$, $\mu^h_{l,m,n}$ of velocity and magnetic field as

$$\frac{d\mu^v_{l,m,n}}{dt}=(l-1)\mu_{l-1,m+1,n}+\alpha_2n(n-1)\mu_{l,m,n-2}+m^2\mu_{l,m,n}-2-l(l-1)\mu_{l-2,m+1,n+1}+eml(l-1)\mu_{l-2,m,n}$$
$$-r_1m\mu_{l,m,n}-\beta\mu_{l+1,m-1,n}-\alpha_1b(t)lm\mu_{l-1,m-1,n}+\alpha_1a(t)m(n-1)\mu_{l,m-2,n} \quad (3\text{-}103)$$

$$\frac{d\mu^h_{l,m,n}}{dt}=(l-1)\mu_{l-1,m+1,n}-km\mu_{l+1,m-1,n}-a_1d(t)lm\mu_{l-1,m-1,n}+a_1c(t)m(m-1)\mu_{l,m-2,n}+nl\mu_{l-1,m+1,n}$$
$$+\varepsilon nl(l-1)\mu_{l-2,m,n-1}-r_2n\mu_{l,m,n}+a_2c(t)n(n-1)\mu_{l,m,n-2} \quad (3\text{-}104)$$

Here $\frac{d}{dt}\mu^v_{l,m,n}=\int_{-\infty}^{+\infty}dx\int_{-\infty}^{+\infty}dv\int_{-\infty}^{+\infty}dhP(x,v,h)x^lv^mh^n$. The kurtosis for displacement and velocity are, respectively, given by

$$K_v=<v^4>/3<v^2>^2-1\ ,\ K_h=<h^4>/3<h^2>^2-1\cdot \quad (3\text{-}105)$$

We get the correlation coefficient as

$$\rho_{v,h}=<(v-<v>)\ h-<h>)>/\sigma_v\sigma_h\cdot \quad (3\text{-}106)$$

Here we assume that a passive particle is initially started at $x=x_0$, $v=v_0$ and $h=h_0$ at time $t=0$. The parameter $\sigma_v$ and $\sigma_h$ are the root-mean-squared value for velocity and magnetic field of joint probability density, respectively.

**Table 5.** Approximate values of the kurtosis, the correlation coefficient, and the moment $\mu_{2,2,0}$, $\mu_{2,0,2}$, $\mu_{0,2,2}$ of velocity and magnetic field in three-time domains, where a passive particle is initially started at $x=x_0, v=v_0, h=h_0$ at time $t=0$.

| time domains | v, h | variables | $K_x$, $K_v$, $K_h$ | $\rho_{x,v}$, $\rho_{x,h}$, $\rho_{v,h}$ | $\mu_{2,2,0}$, $\mu_{2,0,2}$, $\mu_{0,2,2}$ |
|---|---|---|---|---|---|
| $t<<\tau$ | $v$ | $x$ | $K_x=\frac{\tau^2x_0^4}{\alpha_1^2}t^{-8}+\frac{\tau x_0^2}{\alpha_1}t^{-4}$ | $\rho_{x,v}=\frac{\tau x_0v_0}{\alpha_1}t^{-7/2}$ | $\mu^v_{2,2,0}=\frac{\alpha_1^2}{\tau^2}t^6$ |
| | | $v$ | $K_v=\frac{\tau^2v_0^4}{\alpha_1^2}t^{-6}+\frac{\tau v_0^2}{\alpha_1}t^{-3}$ | $\rho_{x,h}=\frac{\tau x_0h_0}{(\alpha_1\alpha_2)^{1/2}}t^{-3}$ | $\mu^h_{2,2,0}=\frac{a_1^3}{\tau^3}t^6$ |
| | | $h$ | $K_h = \frac{\tau^2 h_0^4}{\alpha_2^2} t^{-6} + \frac{\tau h_0^2}{\alpha_2} t^{-3}$ | $\rho_{v,h} = \frac{\tau x_0 h_0}{(\alpha_1 \alpha_2)^{1/2}} t^{-7/2}$ | $\mu^v_{2,0,2} = \frac{\alpha_1 \alpha_2}{\tau^2} t^7$ |
| | | $x$ | $K_x = \frac{\tau^2 x_0^4}{a_1^2} t^{-8} + \frac{\tau x_0^2}{a_1} t^{-4}$ | $\rho_{x,v} = \frac{\tau x_0 v_0}{a_1} t^{-5/2}$ | $\mu^h_{2,0,2} = \frac{a_1^2 a_2}{\tau^3} t^6$ |
| | $h$ | $v$ | $K_v = \frac{\tau^2 v_0^4}{a_1^2} t^{-2} + \frac{\tau v_0^2}{a_1} t^{-1}$ | $\rho_{x,h} = \frac{\tau x_0 h_0}{(a_1 a_2)^{1/2}} t^{-3}$ | $\mu^v_{0,2,2} = \frac{\alpha_1 \alpha_2}{\tau^2} t^5$ |
| | | $h$ | $K_h = \frac{\tau^2 h_0^4}{a_2^2} t^{-4} + \frac{\tau h_0^2}{a_2} t^{-2}$ | $\rho_{v,h} = \frac{\tau v_0 h_0}{(a_1 a_2)^{1/2} \varepsilon} t^{-3/2}$ | $\mu^h_{0,2,2} = \frac{a_1^2 a_2}{\tau^3} t^5$ |
| | | $x$ | $K_x = \frac{x_0^4}{\alpha_1^2} t^{-2} + \frac{x_0^2}{\alpha_1} t^{-1}$ | $\rho_{x,v} = \frac{x_0 v_0}{\alpha_1 \beta} t^{-2}$ | $\mu^v_{2,2,0} = \frac{e \alpha_1 \beta^2}{(r_1 - 2)} t^3$ |
| | $v$ | $v$ | $K_v = \frac{v_0^4}{\alpha_1^2 \beta^4} t^{-2} + \frac{v_0^2}{\alpha_1 \beta^2} t^{-1}$ | $\rho_{x,h} = \frac{\tau x_0 h_0}{(\alpha_1 \alpha_2)^{1/2} \beta} t^{-2}$ | $\mu^h_{2,2,0} = a_1^2 t^4$ |
| | | $h$ | $K_h = \frac{\tau^2 h_0^4}{\alpha_2^2} t^{-2} + \frac{\tau h_0^2}{\alpha_2} t^{-1}$ | $\rho_{v,h} = \frac{v_0 h_0}{(\alpha_1 \alpha_2)^{1/2}} t^{-2}$ | $\mu^v_{2,0,2} = \alpha_1 \alpha_2 t^2$ |
| $t >> \tau$ | | $x$ | $K_x = \frac{x_0^4}{a_1^2 \tau^2} t^{-8} + \frac{x_0^2}{a_1 \tau} t^{-4}$ | $\rho_{x,v} = \frac{\tau x_0 v_0}{(a_1 a_2 \tau)^{1/2} \varepsilon} t^{-5/2}$ | $\mu^h_{2,0,2} = \frac{a_1 a_2}{r_2} t^3$ |
| | $h$ | $v$ | $K_v = \frac{v_0^4}{a_1^2 \varepsilon^4} t^{-6} + \frac{v_0^2}{a_1 \varepsilon^2} t^{-3}$ | $\rho_{x,h} = \frac{x_0 h_0}{(a_1 a_2 \tau)^{1/2}} t^{-2}$ | $\mu^v_{0,2,2} = \frac{\alpha_1 \alpha_2 \beta^2}{(r_1 - 2)\tau} t^3$ |
| | | $h$ | $K_h = \frac{h_0^4}{a_2^2} t^{-4} + \frac{h_0^2}{a_2} t^{-2}$ | $\rho_{v,h} = \frac{v_0 h_0}{(a_1 a_2)^{1/2} \varepsilon} t^{-5/2}$ | $\mu^h_{0,2,2} = \frac{a_1 a_2}{r_2 \tau} t^2$ |
| | | $x$ | $K_x = \frac{x_0^4}{\alpha_1^2 r_1^2} t^{-8} + \frac{x_0^2}{\alpha_1 r_1} t^{-4}$ | $\rho_{x,v} = \frac{x_0 v_0}{\alpha_1 r_1^{1/2}} t^{-5/2}$ | $\mu^v_{2,2,0} = \frac{\alpha_1^2 r_1}{(r_1 - 2)} t^5$ |
| | $v$ | $v$ | $K_v = \frac{v_0^4}{\alpha_1^2} t^{-2} + \frac{v_0^2}{\alpha_1} t^{-1}$ | $\rho_{x,h} = \frac{\tau x_0 h_0}{(\alpha_1 \alpha_2)^{1/2}} t^{-3/2}$ | $\mu^h_{2,2,0} = \frac{a_1^3}{\tau} t^6$ |
| | | $h$ | $K_h = \frac{h_0^4}{\alpha_2^2} t^{-4} + \frac{h_0^2}{\alpha_2} t^{-2}$ | $\rho_{v,h} = \frac{v_0 h_0}{(\alpha_1 \alpha_2)^{1/2} r_2^{1/2}} t^{-3/2}$ | $\mu^v_{2,0,2} = \frac{\alpha_1 \alpha_2 r_1}{\tau} t^6$ |
| $\tau = 0$ | | $x$ | $K_x = \frac{x_0^4}{a_1^2} t^{-6} + \frac{x_0^2}{a_1} t^{-3}$ | $\rho_{x,v} = \frac{x_0 v_0}{a_1} t^{-2}$ | $\mu^h_{2,0,2} = \frac{a_1 a_2^2}{r_2} t^3$ |
| | $h$ | $v$ | $K_v = \frac{v_0^4}{a_1^2} t^{-2} + \frac{v_0^2}{a_1} t^{-1}$ | $\rho_{x,h} = \frac{x_0 h_0}{a_1} t^{-2}$ | $\mu^v_{0,2,2} = \frac{\alpha_1 \alpha_2}{(r_1 - 2)} t^3$ |
| | | $h$ | $K_h = \frac{h_0^4}{a_2^2} t^{-2} + \frac{h_0^2}{a_2} t^{-1}$ | $\rho_{v,h} = \frac{v_0 h_0}{(a_1 a_2)^{1/2}} t^{-2}$ | $\mu^h_{0,2,2} = \frac{a_1^2 a_2}{r_2 \tau} t^3$ |

## 4. Summary

In summary, we firstly have derived the Fokker-Planck equation for a passive particle with harmonic and viscous force, perturbative force, and no harmonic and viscous force. Secondly, we have derived the Fokker-Planck equation in an incompressible conducting fluid of magnetic field. We approximately obtain the solution of the joint probability density by using double Fourier transforms.

The mean squared displacement and the mean squared velocity for joint probability density of $x$ and $v$ is given by Eq. (2-20) and Eq. (2-29) in $t << \tau$ (Eq. (2-46) and Eq. (2-47) in $t >> \tau$, and Eq. (51) in $\tau = 0$). The mean squared displacement and mean squared velocity for a passive particle with both harmonic and viscous force and perturbative force behavior like the super-diffusion with $\sim t^4$ in $\tau = 0$, while those to $\sim t^3$ in $t >> \tau$. The mean squared velocity for a particle with no harmonic and viscous forces has the Gaussian form, which is the normal diffusion with time $< v^2(t) > \sim t$ in $\tau = 0$. The moment behaviors as $\mu_{2,2} \sim t^5$ in $t << \tau$, $t >> \tau$ and $\tau = 0$, not consistent with our result. With perturbative force $+cx^3$, the moment becomes $\mu_{2,2} \sim t^6$ in three-time domains, not consistent with our result. The moment $\mu_{2,2}$ for joint probability density with no harmonic and viscous forces $-\beta x - rv$ particularly scales to time $\sim t^6$ in $t << \tau$, consistent with our result. Other values $\mu_{m,n}$ will be published elsewhere. In addition, the approximately calculated values for the kurtosis, correlation coefficient, and moment from moment equation are provided in Table 1 (the case with harmonic and viscous forces), Table 2 (the case with perturbative force), and Table 3 (the case with no harmonic and viscous forces). In Table 4, we compare values of the joint probability density, mean squared displacement, and mean squared velocity for a passive particle with harmonic and viscous forces $-\beta x - rv$, perturbative force $+cx^3$ and no harmonic and viscous forces $-\beta x - rv$.

In the hydro-magnetic case of two variables, i.e. velocity and time-dependent magnetic field, the mean squared velocity for joint probability density of velocity and magnetic field has super-diffusive form with time $\sim t^3$ in $t >> \tau$, while mean squared displacement for joint probability density of velocity and magnetic field reduces to be time $\sim t^4$ in $t << \tau$. The solution in $t >> \tau$ and $\tau = 0$ behaviors normal diffusion with mean squared magnetic field $< h^2(t) > \sim t$. In short-time domain $t << \tau$, the moment for an incompressible conducting fluid of magnetic field becomes super-diffusion with $\mu^h_{2,0,2} \sim t^6$, consistent with our result. Particularly, in $\tau = 0$, $\mu^v_{2,2,0}$ ($\mu^v_{2,0,2}$, $\mu^v_{0,2,2}$) scales to $\sim t^5$ ($\sim t^6$, $\sim t^3$), consistent with our result. Table 5 summarize the result of calculating kurtosis, correlation coefficient, and moment of probability density from the equations of two variables, velocity and magnetic field, following the initial state $x = x_0$, $v = v_0$, $h = h_0$. Other values will be published elsewhere.

Despite much interest, the exact solutions for distributions of higher-order processes are rare, but the approximate solution of the probability distribution function is found in this paper. The approximate solution of the Navier-Stokes equation has not been solved yet, which is a simple approximate solution from our model. It is hoped in future that we extend our model to the generalized Langevin equation or the equation of motion with other forces. The results obtained can be compared and analyzed with other theories, computer-simulations, and experiments [34-39]. The other results will be continuously published in other journals.

**Acknowledgments**

This research was financially supported by the Ministry of Small and Medium-sized Enterprises (SMEs) and Startups (MSS), Korea, under the "Startup Growth Technology Development Project (R&D, 1425153351)" supervised by the Korea Technology and Information Promotion Agency for SMEs.